\documentclass[letter]{aa}

\usepackage{graphicx}
\usepackage{txfonts}
\usepackage{hyperref}
\usepackage{xcolor}
\hypersetup{
colorlinks,
linkcolor={red!80!black},
citecolor={blue!80!black},
urlcolor={blue!80!black}
}

\newenvironment{myfont}{\fontfamily{ppl}\selectfont}{\par}
\DeclareTextFontCommand{\textmyfont}{\myfont}

\newcommand{\code}[1]{\texttt{#1}}

\def\nifs{\iso{56}Ni}

\def\cm3{cm$^{-3}$}
\def\kms{\mbox{km~s$^{-1}$}}
\def\msunyr{$M_{\odot}$\,yr$^{-1}$}

\def\msun{$M_{\odot}$}

\def\one{\ts {\,\sc i}}
\def\two{\ts {\,\sc ii}}
\def\three{\ts {\,\sc iii}}

\def\beq{\begin{equation}}
\def\eeq{\end{equation}}

\def\lesssim{\mathrel{\hbox{\rlap{\hbox{\lower4pt\hbox{$\sim$}}}\hbox{$<$}}}}
\def\gtrsim{\mathrel{\hbox{\rlap{\hbox{\lower4pt\hbox{$\sim$}}}\hbox{$>$}}}}

\def\one{{\,\sc i}}
\def\two{{\,\sc ii}}
\def\three{{\,\sc iii}}

\def\v1d{{\code{V1D}}}

\def\cmfgen{{\code{CMFGEN}}}
\def\heracles{{\code{HERACLES}}}

\def\ergs{erg\,s$^{-1}$}

\def\oidoub{[O\one]\,$\lambda\lambda$\,$6300,\,6364$}

\def\mgiidoub{Mg\two\,$\lambda\lambda$\,$2795,\,2802$}

\newcommand{\iso}[2]{\ensuremath{^{#1}\rm{#2}}}

\begin{document}

\title{Modeling the signatures of interaction in Type II supernovae: \\
UV emission, high-velocity features, broad-boxy profiles}

\titlerunning{Signatures of interaction in Type II supernovae}

\author{Luc Dessart\inst{\ref{inst1}}
  \and
   D. John Hillier\inst{\ref{inst2}}
  }

\institute{
Institut d'Astrophysique de Paris, CNRS-Sorbonne Universit\'e, 98 bis boulevard Arago, F-75014 Paris, France.\label{inst1}
\and
    Department of Physics and Astronomy \& Pittsburgh Particle Physics,
    Astrophysics, and Cosmology Center (PITT PACC),  \hfill \\ University of Pittsburgh,
    3941 O'Hara Street, Pittsburgh, PA 15260, USA.\label{inst2}
  }

   \date{}

  \abstract{
  Because mass loss is a fundamental phenomenon in massive stars, interaction with circumstellar material (CSM) should be universal in core-collapse supernovae (SNe). Leaving aside the extreme CSM density, extent, or mass typically encountered in Type IIn SNe, we investigate the diverse long-term radiative signatures of interaction between a Type II SN ejecta and CSM corresponding to mass loss rates up to 10$^{-3}$\,\msunyr. Because these CSM are relatively tenuous and optically-thin to electron-scattering beyond a few stellar radii, radiation hydrodynamics is not essential and one may treat the interaction directly as an additional power source in the non-local thermodynamic equilibrium radiative transfer problem. The CSM accumulated since shock breakout forms a dense shell in the outer ejecta and leads to high-velocity absorption features in spectral lines, even for negligible shock power. Besides Balmer lines, such features may appear in Na\one\,D, He\one\ lines etc. A stronger interaction strengthens the continuum flux (preferentially in the UV), quenches the absorption of P-Cygni profiles, boosts the \mgiidoub\ doublet, and fosters the production of a broad boxy H$\alpha$ emission component. The rise in ionization in the outer ejecta may quench some lines (e.g., the Ca\two\ near-infrared triplet). The interaction power emerges preferentially in the UV, in particular at later times, shifting the optical color to the blue, but increasing modestly the optical luminosity.  Strong thermalization and clumping seem to be required to make an interaction superluminous in the optical. The UV range contains essential signatures that provide critical  constraints to infer the mass loss history and inner workings of core-collapse SN progenitors at death.
  }

   \maketitle
%

\section{Introduction}

Mass loss is a fundamental phenomenon affecting the evolution of massive stars. Steady-state mass loss in the form of a wind can significantly reduce the mass of a star and affect many of its properties throughout its life \citep{maeder_meynet_87,langer_massive_94}, as well as the mass of the metal-rich core, the final yields, and the compactness at collapse \citep{whw02,renzo_mdot_17}. Steady or dynamical mass transfer through Roche-lobe overflow in interacting binaries are typically short lived but manyfold stronger than wind mass loss and can thus lead to a dramatic change in the evolution of a massive star \citep{podsiadlowski_92,wellstein_langer_99,eldridge_08_bin,langer_araa_mdot_12}. Interacting binaries are likely central for the production of stripped-envelope supernovae \citep[SNe;][]{yoon_ibc_10}. Hence, stellar wind mass loss and mass transfer in interacting binaries play a central role in determining the gross properties of the SN progenitor and the SN type.

However, the radiative properties of SNe depend also on the complex internal structure of the star at death and the  structure of its close environment. The final structure of the star affects the shock propagation through the envelope prior to shock breakout.
When the close environment of the star is not a vacuum, as generally assumed for simplicity, an interaction takes place with the circumstellar material (CSM). While the existence of CSM is not surprising for massive stars, a number of observed SNe suggest that this CSM may be exceptionally dense, in the sense that it would correspond to wind mass loss rates manyfold greater than standard steady-state mass loss rates typically inferred in massive stars (see, for example, \citealt{dejager_mdot_88}). This CSM may also extend out to large distances above the star surface, from several 10$^{14}$\,cm  \citep[SN\,2013fs;][]{d17_13fs,moriya_13fs_17,morozova_2l_2p_17,yaron_13fs_17}, to 10$^{15}$\,cm \citep[SN\,2020tlf;][]{wynn_20tlf_22}, to several 10$^{15}$\,cm \citep[SN\,1998S;][]{leonard_98S_00,fassia_98S_01,chugai_98S_01,D16_2n}, and up to 10$^{16}$\,cm \citep[SN\,2010jl;][]{zhang_10jl_12,fransson_10jl,D15_2n}. While large relative to the progenitor radius, these distances correspond to mass loss episodes occurring on year time scales before core collapse and thus may point to dynamical phenomena tied to the last stages of massive star evolution \citep{quataert_shiode_12,fuller_rsg_17}.

A critical spectroscopic signature of interaction in a SN is the presence of narrow and symmetric emission line profiles instead of the expected Doppler-broadened P-Cygni profiles \citep{schlegel_iin_90,SS_88Z_91,chugai_98S_01,d09_94w}. This is most obvious in  H\one\ Balmer lines of Type II SNe (e.g., SN\,1998S), but also seen in lines of He\one, He\two, or more highly ionized ions of C, N, or O (e.g., SN\,2013fs; \citealt{yaron_13fs_17}). Such lines imply a reprocessing of the radiation from the underlying SN by the unshocked, quasistatic, optically-thick CSM.  This CSM releases the SN radiation on a diffusion timescale, broadening significantly the shock breakout signal. In the interaction between ejecta and CSM, the ejecta deceleration implies a transformation of kinetic energy into radiative energy, which boosts the SN luminosity. The CSM also induces a shift of the spectral energy distribution to longer wavelength because the radiation is trapped within a larger optically-thick volume. Together, these effects tend to boost the optical luminosity of the SN.

While many studies have been devoted to superluminous interacting SNe, the impact of mass loss is likely much more universal in core-collapse SNe. For wind mass loss rates stronger than standard but too low to produce an optically-thick CSM, the SN spectra cannot exhibit the narrow spectral signatures that would flag the event as a Type IIn or Type Ibn. However, the shock power may nonetheless be substantial and even supersede the `default' SN luminosity (i.e., in the absence of interaction). Overluminous SNe II exhibiting non standard broad line profiles (e.g., H$\alpha$) are indeed observed (Pessi et al., in prep). For a SN shock ramming at velocity $V_{\rm sh}$ into a steady-state wind with mass loss rate $\dot{M}$ and velocity $V_\infty$, the instantaneous power released by the interaction is
$L_{\rm sh} = \dot{M} V_{\rm sh}^3 / 2 V_\infty
= 3.15 \times 10^{40} \,\,\, \dot{M}_{-5} V_{\rm sh,4}^3 / V_{\infty,2} \,\,\, {\rm erg}\,{\rm s}^{-1} \, ,
$
where $\dot{M}_{-5} \equiv \dot{M} / 10^{-5} M_\odot$ yr$^{-1}$, $V_{\rm sh,4} \equiv V_{\rm sh}$/10000\,km\,s$^{-1}$ and $V_{\infty,2} \equiv V_\infty/100$\,km\,s$^{-1}$. If in addition $V_{\rm sh}$ is fixed, the shock power is constant in time. The representative luminosity of a Type II SN during the photospheric phase, which is a few times \,10$^{42}$\,erg\,s$^{-1}$ \citep{bersten_lbol_09}, can be rivaled by the power from the ejecta interaction with a wind mass loss rate of 10$^{-3}$\,\msunyr. This mass loss rate is still a factor of a 100 weaker than inferred for a superluminous Type IIn SN like 2010jl \citep{fransson_10jl,D15_2n}. Furthermore, if the wind density remains sufficiently high at large distances, the shock power should eventually `win' over the inner ejecta luminosity (powered by radioactive decay). For example, for $\dot{M}_{-5} =$\,3.2, $V_{\rm sh,4}=$\,1.0, and  $V_{\infty,2}=$\,1.0, the interaction power is 10$^{41}$\,\ergs\ and comparable to the power at 300\,d from the decay of 0.1\,\msun\ of \nifs. The contribution from interaction power to the SN radiation at some stage in the evolution of a core-collapse SN is thus likely.

For CSM configurations corresponding to wind mass loss rates of 0.1\,\msunyr\ and extending over large scales, a multi-group radiation hydrodynamics treatment is essential. This approach captures the dynamics of the interaction, the deceleration of the ejecta, and the continuous pile-up of CSM into a dense shell. Furthermore, as the CSM becomes optically thick following the radiative precursor, the radiative transfer problem is time dependent, and so are the gas properties. As long as the shock is embedded in this optically-thick CSM, photons emerge from the unshocked CSM. This situation changes as the shock progresses outwards and escaping photons increasingly originate from the shocked CSM and ejecta. Such configurations correspond to superluminous SNe \citep{moriya_rsg_csm_11,D15_2n}.

In this letter, we leave aside these extreme CSM configurations. Instead, we consider the interaction of a standard Type II SN ejecta with a CSM corresponding to a wind mass loss rate that is always too weak to make the CSM optically thick to electron scattering (except perhaps in the immediate vicinity of the progenitor surface, which is quickly swept up by the ejecta on a day timescale). In this case, there is no need for radiation hydrodynamics since 1) the shock power is known analytically and given above by $L_{\rm sh}$; 2) the CSM is transparent so its contribution in emission and absorption may be neglected; 3) and the CSM is too tenuous to appreciably decelerate the ejecta. Hence, in contrast to the approach used in our previous studies \citep{D15_2n, D16_2n, d17_13fs}, we simulate the interaction between ejecta and CSM directly within \cmfgen\ \citep{HD12}. While this ignores the dynamics of the problem, this allows for a much better treatment of the radiative transfer and the effects associated with departures from local thermodynamic equilibrium (LTE). It also allows for the long-term modeling of an interaction over weeks, months, and even years (see Sect.~\ref{sect_conc}). In the next section, we present our numerical approach, including the treatment of the interaction power in \cmfgen\ and the set of simulations we cover in this work. We then present our results in Sect.~\ref{sect_res}. We conclude in Sect.~\ref{sect_conc}.

 \begin{figure}
\centering
\includegraphics[width=\hsize]{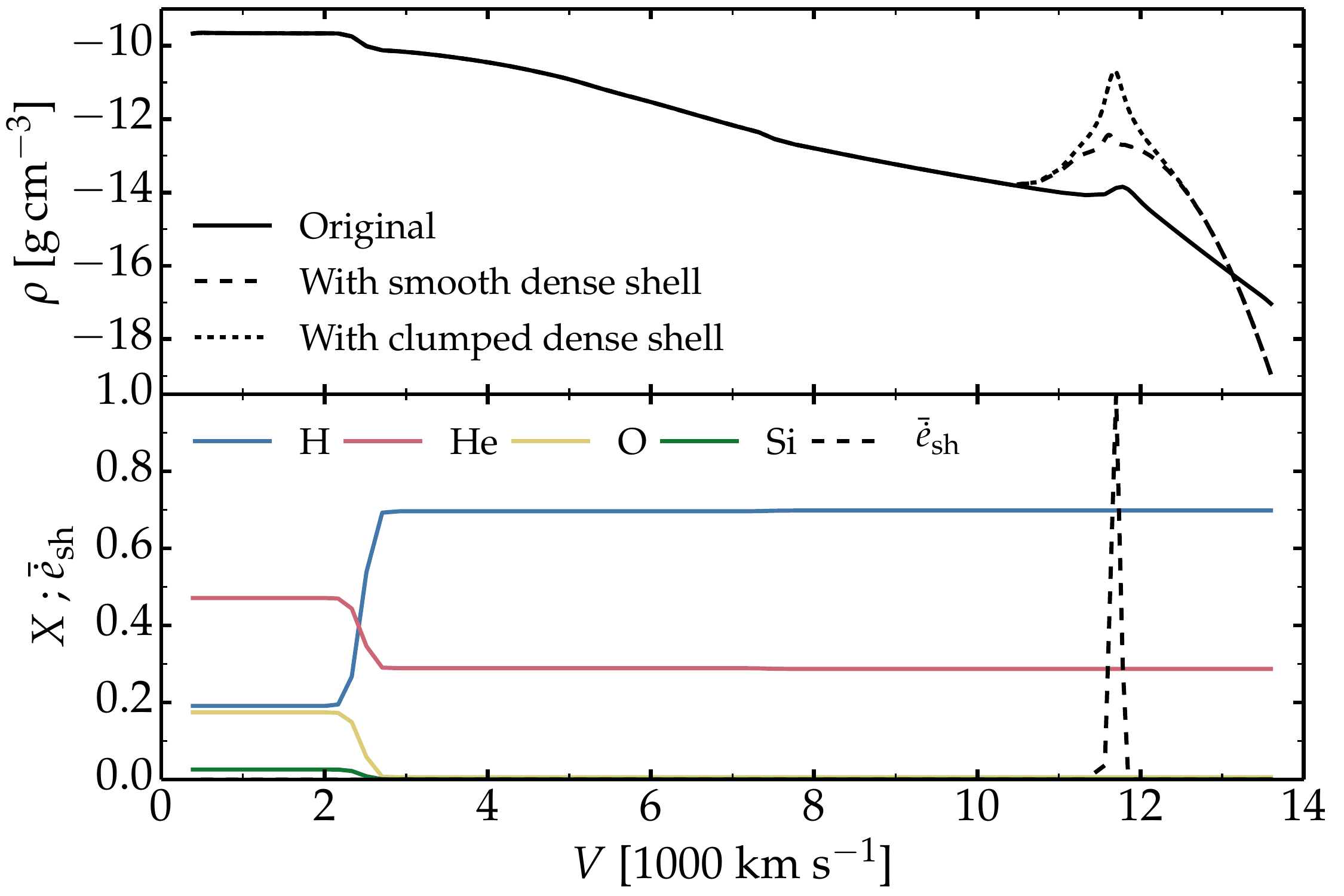}
\vspace{-0.6cm}
\caption{Top: Ejecta density structure versus velocity for the original Type II SN model at 10\,d after explosion (black). For this work, we introduce a dense shell at 11700\,\kms\ (blue), which we clump (maximum compression factor of 100; red). Bottom: Corresponding mass fraction versus velocity for H, He, O, and Si, as well as for the adopted (normalized) shock-power profile associated with the interaction between CSM and ejecta.
 \label{fig_init_prop}
}
\end{figure}

\section{Numerical setup}
\label{sect_setup}

All models presented in this work start with the same default Type II SN ejecta at an age of 10\,d. It corresponds to a star of 15\,\msun\ initially, evolved at solar metallicity, and that exploded to produce an ejecta of 10.81\,\msun\ with a kinetic energy of $1.3 \times 10^{51}$\,erg and 0.03\,\msun\ of \nifs\ (for details on the method, see \citealt{HD19}). To mimic the presence of a cocoon of material in the direct vicinity of the progenitor red-supergiant (RSG) star, something that is likely typical in SNe II-P with super-wind phases prior to explosion as inferred for SN\,2013fs \citep{yaron_13fs_17}, we introduce in the initial density structure a dense shell of 0.1\,\msun\ at a velocity of 11700\,\kms\ (compatible with model predictions; \citealt{d17_13fs}). Because of multidimensional instabilities, this dense shell should break-up in 3D and exhibit strong clumping. We thus spread the dense shell over a characteristic scale set to 10\,\% of the local radius, and introduce clumping to reach a maximum compression factor of 100 (volume filling factor of 1\,\%). Numerically, this allows us to resolve this dense shell in 1D while preserving a high density for the clumped material in the dense shell (Fig.~\ref{fig_init_prop}).

All models are evolved assuming a fixed shock velocity ($V_{\rm sh,4}=$\,1.17) and a steady progenitor wind ($\dot{M}_{-5}$ from zero to 100 and  $V_{\infty,2}=$\,0.5), corresponding to a fixed interaction power from zero to 10$^{43}$\,\ergs. The accumulated swept-up mass by the ejecta after one year is $0.001  \dot{M}_{-5} V_{\rm sh,4} / V_{\infty,2} M_\odot$ (i.e., about 0.1\,\msun\ for $\dot{M}_{-5}=$\,100). Being relatively small, we decide to neglect it and go further and neglect any dynamical influence of the CSM on the initial ejecta density structure. We also neglect the absorption and emission from such a low-density CSM. We reset its properties so that its density is negligible and its velocity is homologous. With these adjustments we can model the SN ejecta and the interaction using the same approach as we use to model non-interacting SN ejecta in \cmfgen.

\begin{figure}
\centering
\includegraphics[width=\hsize]{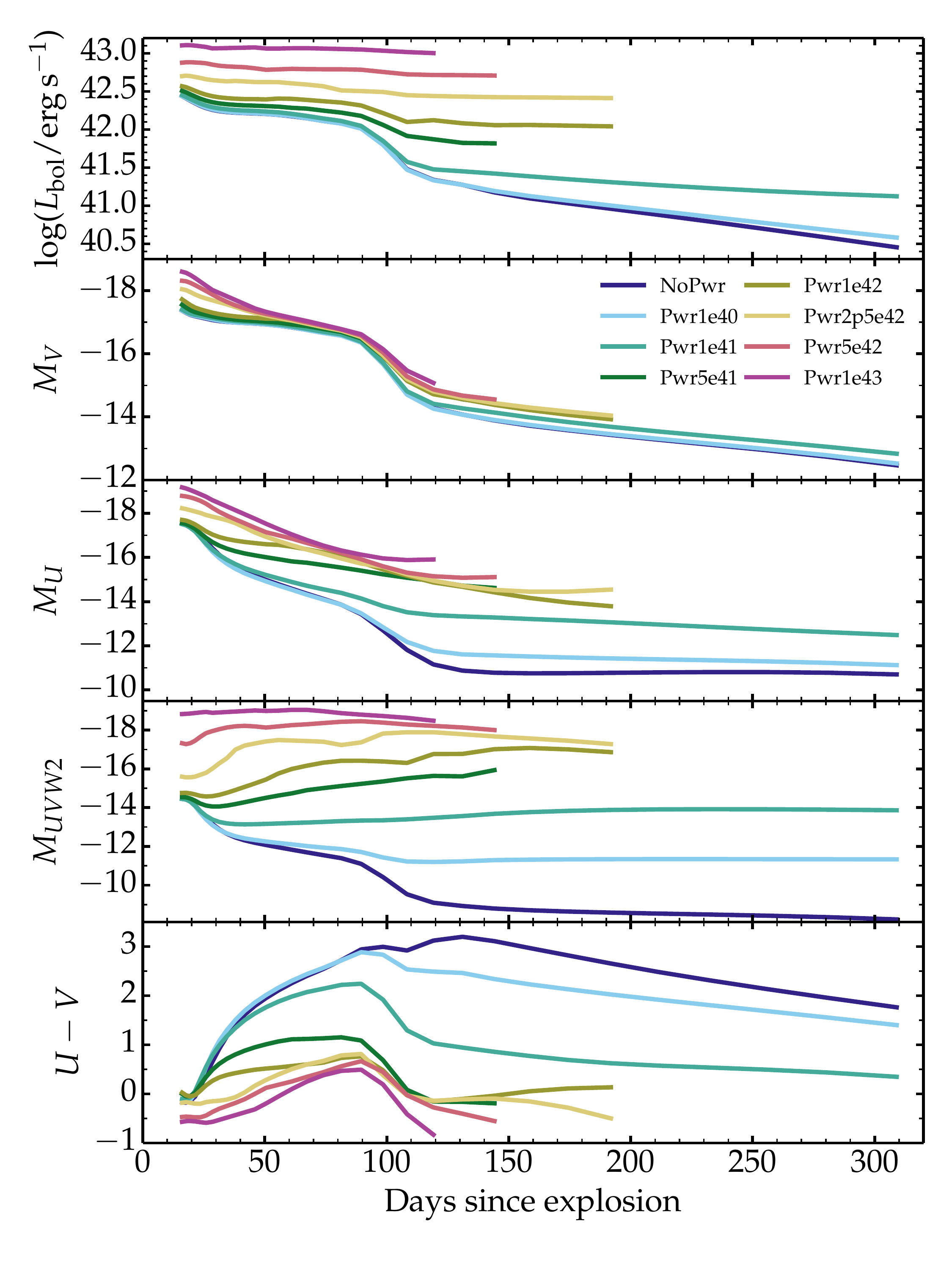}
\vspace{-1cm}
\caption{Photometric properties for our set of models. From top to bottom, we show the evolution of the bolometric luminosity, $M_V$, $M_U$, $M_{UVW2}$, and the $U-V$ color. \label{fig_lbol}
}
\end{figure}

How one should deposit the interaction power is a delicate matter. In radiation-hydrodynamics simulations of  interactions, power is injected in the forward and in the reverse shocks that bound the dense shell. Because of the break-up of the dense shell in 3D, which leads to clumping and turbulence etc, the deposition of shock power is highly complex. Here, we focus on that part of the shock power that is absorbed in the ejecta and ignore any high-energy radiation that escapes. Hence the power that we deposit underestimates the true power of the shock (particularly for the forward shock, but much less so for the reverse shock). As a first step, we deposit the shock power in the dense shell and we model how it thermalizes in that region. In practice, we adopt a Gaussian profile for the deposition of shock power, with a centroid at the center of the dense shell at 11700\,\kms\ with a characteristic width of 75\,\kms\ (dashed-curve in the bottom panel of Fig.~\ref{fig_init_prop}; the clumping profile takes the same form).  We treat equivalently shock power and decay power in \cmfgen, that is, the shock power is injected as high-energy radiation and degraded in the same way as $\gamma$-rays from radioactive decay. Clumping is treated as in \citet{d18_fcl} and we therefore neglect porosity. Nonthermal effects are thus computed consistently by the nonLTE solver.  Because of the rapid change in density, ionization, and temperature within the dense shell, we enforce a high resolution in this region. Regridding is also performed in individual models to resolve accurately the recombination fronts.

In this work, we present time sequences for a SN II ejecta under the influence of eight different interaction powers (but constant in time for any given model sequence). These seven distinct sequences take shock powers of 10$^{40}$, 10$^{41}$, 5 $\times$ 10$^{41}$, 10$^{42}$, 2.5 $\times$ 10$^{42}$, 5 $\times$ 10$^{42}$, and 10$^{43}$\,\ergs\ (the corresponding model names are Pwr1e40 etc). One sequence without interaction power serves as a reference (model NoPwr). All models have the same density structure at any given SN age and were evolved until 150--300\,d -- the models change little once nebular.  The simulations are  time dependent and contain all the assets of the standard simulations performed in the past with \cmfgen\ (for similar but noninteracting SNe II models, see, for example, \citealt{HD19}). Different ejecta models, levels of clumping (which is known to impact the temperature and the gas ionization), shock velocities, progenitor mass loss history, or composition will be considered in future studies.

\section{Results}
\label{sect_res}

The top panel of Fig.~\ref{fig_lbol} shows the bolometric light curves for our set of models.  For increasing interaction power, the luminosity from the underlying ejecta (stemming from energy deposited by the SN shock prior to shock breakout as well as from the continuous power supply from radioactive decay) is progressively dominated by the interaction power and eventually swamped for $L_{\rm sh} \gtrsim 5 \times 10^{42}$\,\ergs\ during the photospheric phase. Although the models span three orders of magnitude in interaction power, there is only a factor of 5.6 difference in bolometric luminosity at 40\,d after explosion. This arises from the large intrinsic luminosity of noninteracting SN II ejecta during the photospheric phase, which stems from the release of stored radiation energy. Because the SN ejecta is typically ten times fainter at nebular times, an interaction power of $\gtrsim$\,10$^{41}$\,\ergs\ is more easily discerned after about 120\,d.

\begin{figure*}
\centering
\includegraphics[width=\hsize]{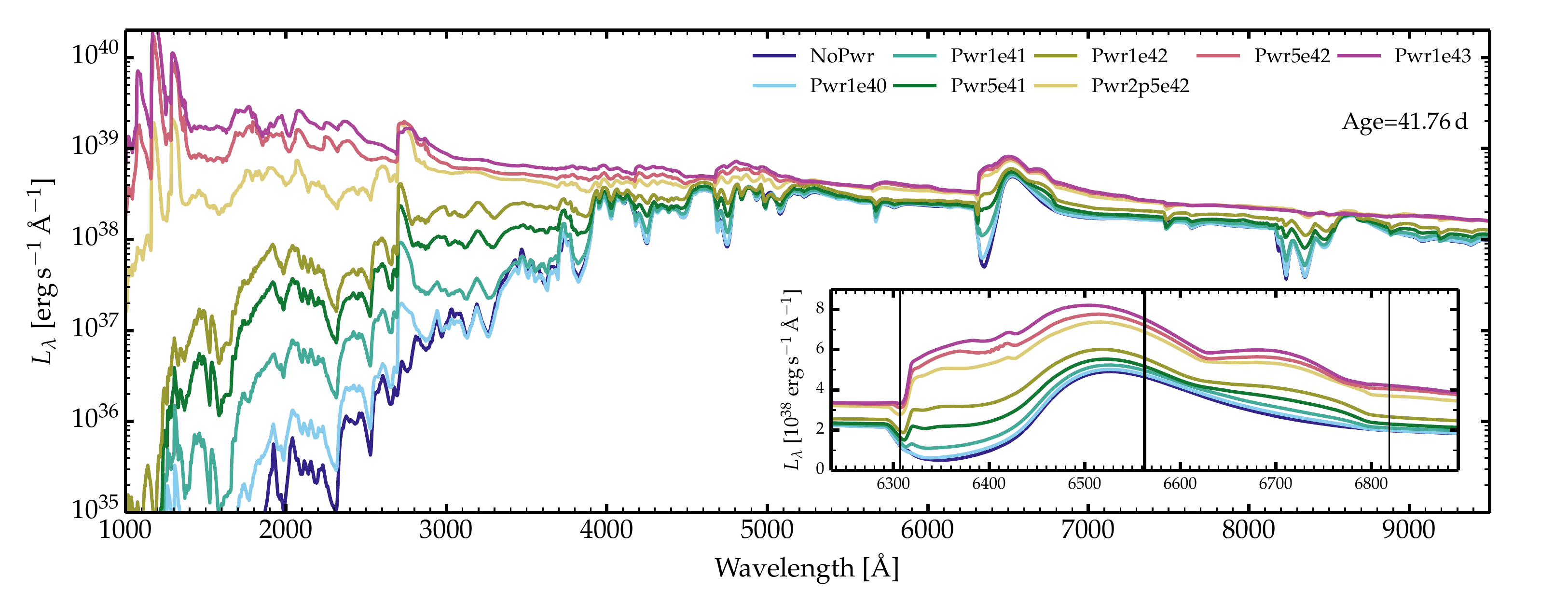}
\vspace{-0.75cm}
\caption{Comparison of the observer's frame luminosity in the UV and optical ranges for our set of models with an interaction power covering from zero to 10$^{43}$\,\ergs\ and at a time of 41.76\,d. The inset zooms on the H$\alpha$ region. The thick vertical line represents the rest wavelength of H$\alpha$, and the thin vertical lines indicate the wavelength at $\pm$\,11700\,\kms\ away from that rest wavelength.
\label{fig_spec_all}
}
\end{figure*}

\begin{figure*}
\centering
\includegraphics[width=0.49\hsize]{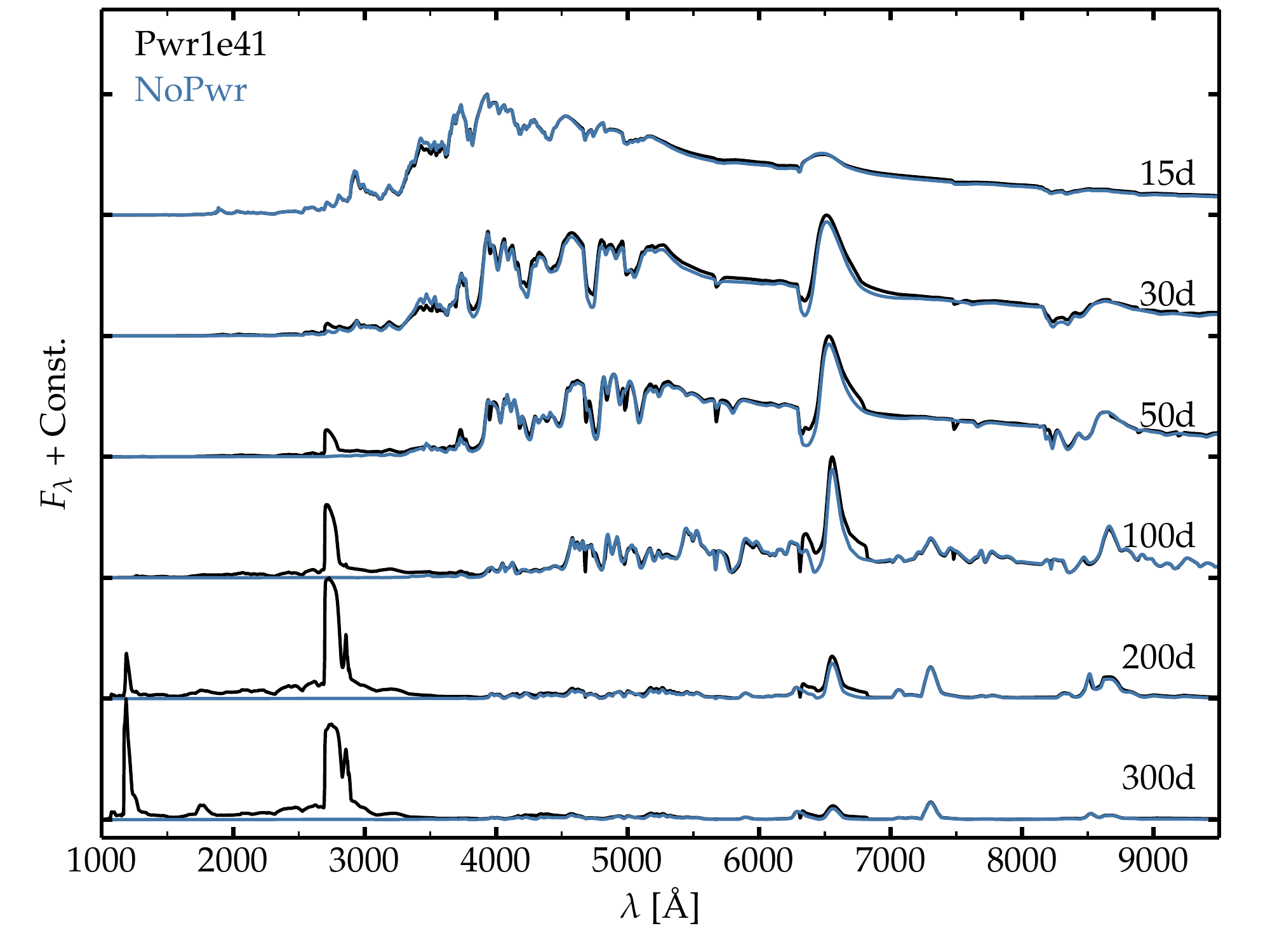}
\includegraphics[width=0.3065\hsize]{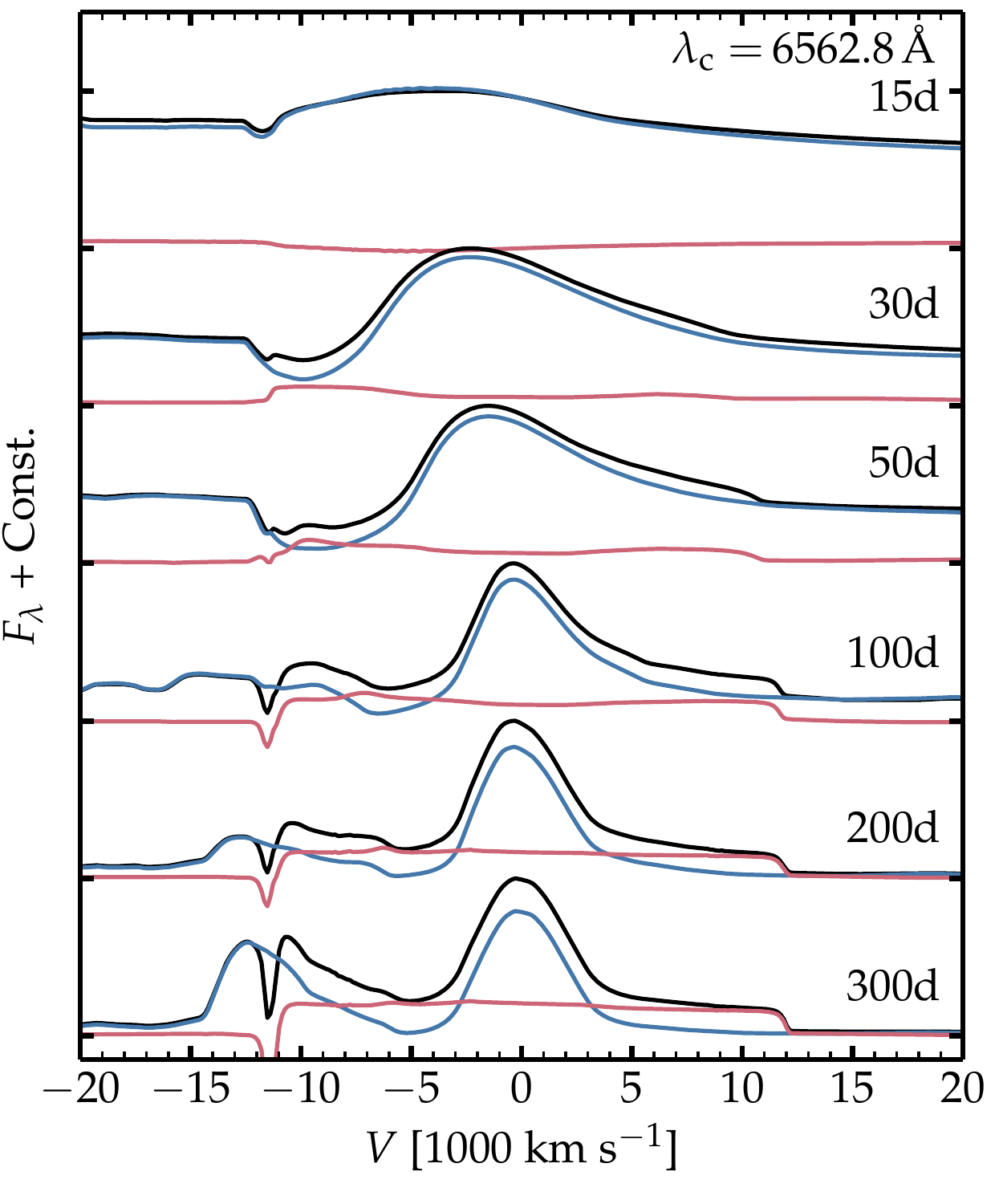}
\vspace{-0.4cm}
\caption{Left: Time evolution of the spectral properties in the UV and in the optical for the SN II model with an interaction power of 10$^{41}$\,erg\,s$^{-1}$ (black) and the model without interaction power (blue). Both models have, however, the same ejecta density structure (short-dashed curve in Fig.~\ref{fig_init_prop}). For each epoch, the spectra are divided by the maximum flux of the brighter model and then stacked for visibility. Right: Same as left but now showing the spectra versus Doppler velocity and centered on the rest wavelength of H$\alpha$. The red curve shows the flux difference between the two models.
\label{fig_pwr2}
}
\end{figure*}

The lower panels of Fig.~\ref{fig_lbol} give some photometric properties. It is essentially impossible from the $V$-band brightness (0.5\,mag offset between models at 42\,d) to guess the presence of an interaction in this parameter space, but the interaction is strikingly apparent in the $U$ band, and more so as we progress to the UV (with offsets of several magnitudes).  The sensitivity in $U$ makes the $U-V$ color a good probe of these interaction models.

Figure~\ref{fig_spec_all} compares the synthetic spectra from 1000 to 9500\,\AA\ at 41.76\,d after explosion (see Fig.~\ref{fig_spec_all_108d} for the counterpart at 108.3\,d). At that time, the maximum offset is only 50\,\% in optical luminosity  but the disparity between models is huge in the UV. Models without interaction power are UV faint as a result of a cold photosphere essentially at the H recombination temperature and affected by strong metal line blanketing. With increasing interaction power, the UV flux progressively rises and the \mgiidoub\ doublet emerges and strengthens. For the models with the highest powers,  other emission lines appear in the far-UV, namely multiplets of Si\three\ and Fe\three\ around 1110\,\AA, Ly$\alpha$, or C\two\ at 1335\,\AA. Hence, interaction power is an efficient process to produce superluminous SNe but such transients are optically superluminous only if the interaction power is strongly degraded into low energy photons. In this context, observations in both UV and optical are essential to establish the bolometric luminosity of a SN and determine the presence of an interaction.

In contrast to the diverse UV landscape sketched by the models, the interaction yields subtle, though compelling, spectral features in the optical. Numerous line profiles including H$\alpha$, H$\beta$, Na\one\,D, or the Ca\two\ near-infrared triplet, exhibit a strong and narrow absorption at the Doppler velocity corresponding to the dense shell (i.e., a `high velocity' or HV feature in SN jargon). This HV notch is present even without interaction power. However,  as interaction power is increased, this absorption becomes more obvious (see inset for H$\alpha$ in Fig.~\ref{fig_spec_all}). For a given model sequence, these HV features emerge at the onset of the recombination epoch (around 30--50\,d) and strengthen thereafter (Fig.~\ref{fig_pwr2}).  Our results confirm the findings of \citet{chugai_hv_07} based on observations of SNe\,1999em and 2004dj -- larger samples of Type II SN spectra reveal that such HV features are not rare although they tend to be more frequently observed in H$\alpha$ and H$\beta$ \citep{gutierrez_pap1_17}.

The HV features arise because the thin shell at large velocity is much larger than the continuum emitting region. These notches are narrow because the shell is narrow and because projection effects are small. Such a shell may have a complex formation history. It first forms at the time of shock breakout, but this alone produces a negligible shell mass. More mass can be accumulated if some dense CSM is present close to the star, as in SN\,2013fs \citep{yaron_13fs_17}, and in addition, or alternatively, if some lower density CSM is present over a large volume, as proposed by \citet{chugai_hv_07}. We do find in our simulations that even without interaction power, a HV feature may be present (model NoPwr shows a HV feature in H$\alpha$, H$\beta$, Na\one\,D, and in the Ca\two\, near infrared triplet at $\gtrsim$\,50\,d).

As interaction power is increased (mass loss rates greater than 10$^{-5}$\,\msunyr, thus barely above the typical RSG mass loss rates), a broad H$\alpha$ emission component appears (right panel in Fig.~\ref{fig_pwr2}). Because of optical depth effects, it is strongly blueshifted early in the photospheric phase but eventually turns into a boxy emission bounded within $\pm$\,11700\,\kms\ (i.e., corresponding to the dense shell velocity). A boxy profile is simply a consequence of having a narrow emitting shell and a Hubble flow.  In model Pwr1e41 at 300\,d, this extended emission overlaps in the blue with \oidoub, making the red component of the O\one\ doublet appear stronger than the blue component (right panel of Fig.~\ref{fig_pwr2}). This boxy emission extends to the red of the H$\alpha$ rest wavelength and makes its origin clear : It only occurs in the presence of interaction (model NoPwr does not have it). Such a profile has been seen at nebular times in SN\,2017ivv \citep{gutierrez_17ivv_20}.

In the models with the highest power, the model spectra in the UV and in the optical strongly depart from the NoPwr model (Fig.~\ref{fig_spec_all}). Besides the huge UV flux, a blue continuum flux weakens the line absorptions in the optical. Some lines are also weakened because the interaction power enhances the ionization and the temperature of the ejecta layers (an ionization wave penetrates the ejecta inwards from the CDS), overionizing certain species (e.g., at 41.76\,d, Ca is everywhere Ca$^{2+}$ in model Pwr1e43 but partially Ca$^+$ in model Pwr1e40). The broad boxy emission component is very strong in model Pwr1e43 during the photospheric phase, with a slight excess near line center reminiscent of the H$\alpha$ emission seen in model NoPwr. An H$\alpha$ profile with a weak or no absorption component testifies for the presence of an interaction \citep{HD19} and the connexion to brighter and faster declining Type II SN light curves \citep{gutierrez_ha_14}. At nebular times, the outer dense shell in the models with the highest power is hot and highly ionized, which quenches the broad boxy H$\alpha$ emission component (Figs.~\ref{fig_pwr4b}--\ref{fig_pwr6}).

In nearly all of our simulations, the interaction power eventually dominates over decay power as we progress in the nebular phase.  The trends seen in a given model with increasing SN age are qualitatively similar to those seen at a given time with increasing interaction power (Fig.~\ref{fig_spec_all}).  In model Pwr1e41 at 300\,d, the offset in flux is primarily confined to the UV (67\,\% of the total flux emerges in the UV) where strong lines have formed (Ly$\alpha$ and \mgiidoub). This strong effect results from our ad-hoc prescription of a fixed interaction power at all times. However, it is clear that the UV range is critical to evaluate the presence of an interaction since the only obvious optical signature is limited to the broad boxy emission component of H$\alpha$, which typically contains of order 1\,\% of the total interaction power injected in the model.

\section{Conclusions}
\label{sect_conc}

In this letter, we have presented a new method for modeling interacting SNe within the time-dependent nonLTE radiative transfer code \cmfgen. The method is designed to cover conditions in which the CSM is optically thin in electron scattering, to complement the alternate configurations with optically-thick CSM handled with \heracles\ and the non-monotonic solver in \cmfgen\ \citep{D15_2n}.
The preliminary simulations presented in this letter were focused on a Type II SN ejecta, modified to exhibit a dense shell at high velocity to mimic the swept-up mass from the interaction with CSM, and influenced by various shock powers. Although it depends on the level of clumping in the dense shell and the efficiency of thermalization therein, the bulk of the interaction power tends to emerge in the UV, making this spectral range essential in this context and emphasizing the need for future observations of transients in the UV \citep{uvex}. In the optical, the presence of a dense shell can give rise to HV features in numerous strong lines (H$\alpha$, Na\one\,D etc), even in the absence of interaction power. However, as the interaction is increased, the HV absorptions tend to strengthen while a broad boxy H$\alpha$ emission component develops.

\begin{figure}
\centering
\includegraphics[width=\hsize]{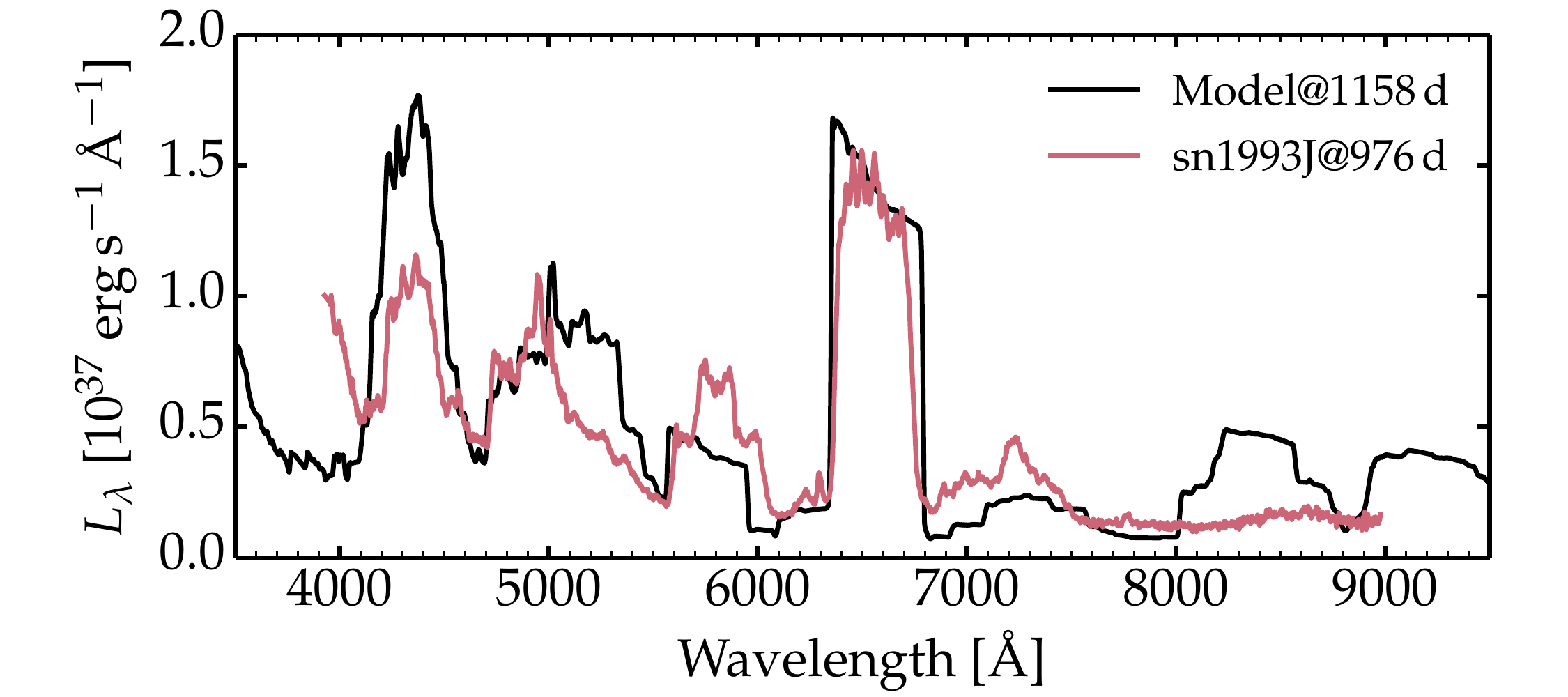}
\vspace{-0.4cm}
\caption{Example calculation for an interaction at a very late time in the nebular phase (black). The ejecta is limited to a 1\,\msun\ dense shell moving at 10000\,\kms\ and powered at a rate of $3 \times 10^{41}$\,\ergs. A uniform and strong clumping corresponding to a volume filling factor of 1\,\% is used to compensate for the spatial spread of the dense shell (set to $\sim$\,10\% of the local radius). This spectrum is reminiscent of the observations of SN\,1993J at a similar epoch \citep[red; ][]{matheson_93j_00a}. The observed spectrum has been corrected for redshift and reddening, and then normalized at 4800\,\AA\ to match the model spectrum.
\label{fig_93J}
}
\end{figure}

The method can be applied to any type of SN ejecta and may help quantify the level at which an interaction with a pre-SN wind can contaminate the SN radiation. Because it is well suited at nebular times, it will allow investigations on the long-term evolution of SN ejecta such as SN\,1993J (an exploratory model is shown in Fig.~\ref{fig_93J}), and help constrain the mass loss history of SN progenitors in the years, decades or centuries prior to explosion.

\begin{acknowledgements}

We acknowledge fruitful discussion with J. Anderson and P. Pessi. This work was supported by the ``Programme National de Physique Stellaire'' of CNRS/INSU co-funded by CEA and CNES. DJH thanks NASA for partial support through the astrophysical theory grant 80NSSC20K0524. This work was granted access to the HPC resources of  CINES under the 2020 allocation A0090410554 made by GENCI, France. This research has made use of NASA's Astrophysics Data System Bibliographic Services.

\end{acknowledgements}


\appendix
\section{Additional figures}

\begin{figure*}
\centering
\includegraphics[width=\hsize]{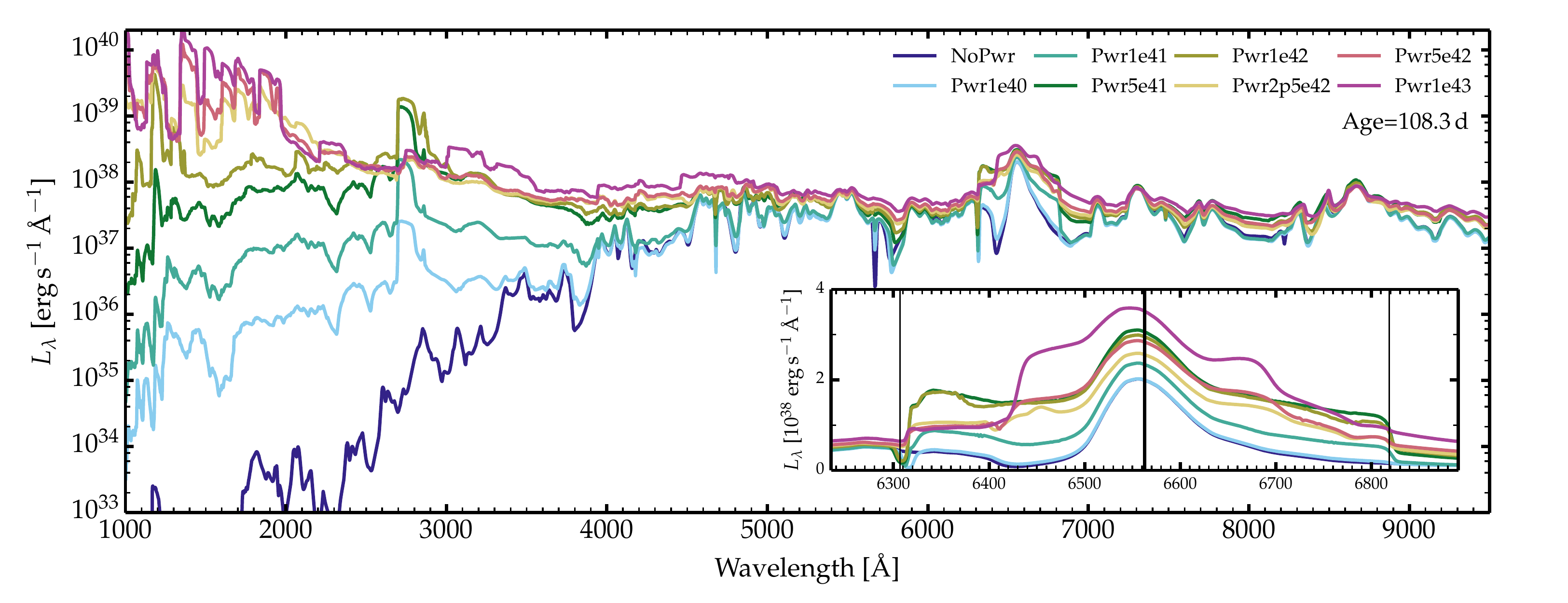}
\vspace{-0.4cm}
\caption{Same as Fig.~\ref{fig_spec_all}, but now at 108.3\,d after explosion.
\label{fig_spec_all_108d}
}
\end{figure*}

In this appendix, we provide additional illustrations to document the results from our radiative transfer simulations.  Figure~\ref{fig_spec_all_108d} shows the UV and optical properties at 108.3\,d. This epoch corresponds to the onset of the nebular phase for all models (see the bolometric light curves in the top panel of Fig.~\ref{fig_lbol}). Compared to the properties at 41.76\,d (Fig.~\ref{fig_spec_all}), the impact of the interaction is greater because the radiation from the noninteracting part of the ejecta is smaller. The lower densities also lead to a greater ionization and temperature, which all conspire to producing a greater boost to the UV luminosity. The broad and boxy emission component in the H$\alpha$ profile is also more developed (except for the highest powers), while the H$\alpha$ emission from the underlying ejecta is weaker and narrower.

\begin{figure*}
\centering
\includegraphics[width=0.61\hsize]{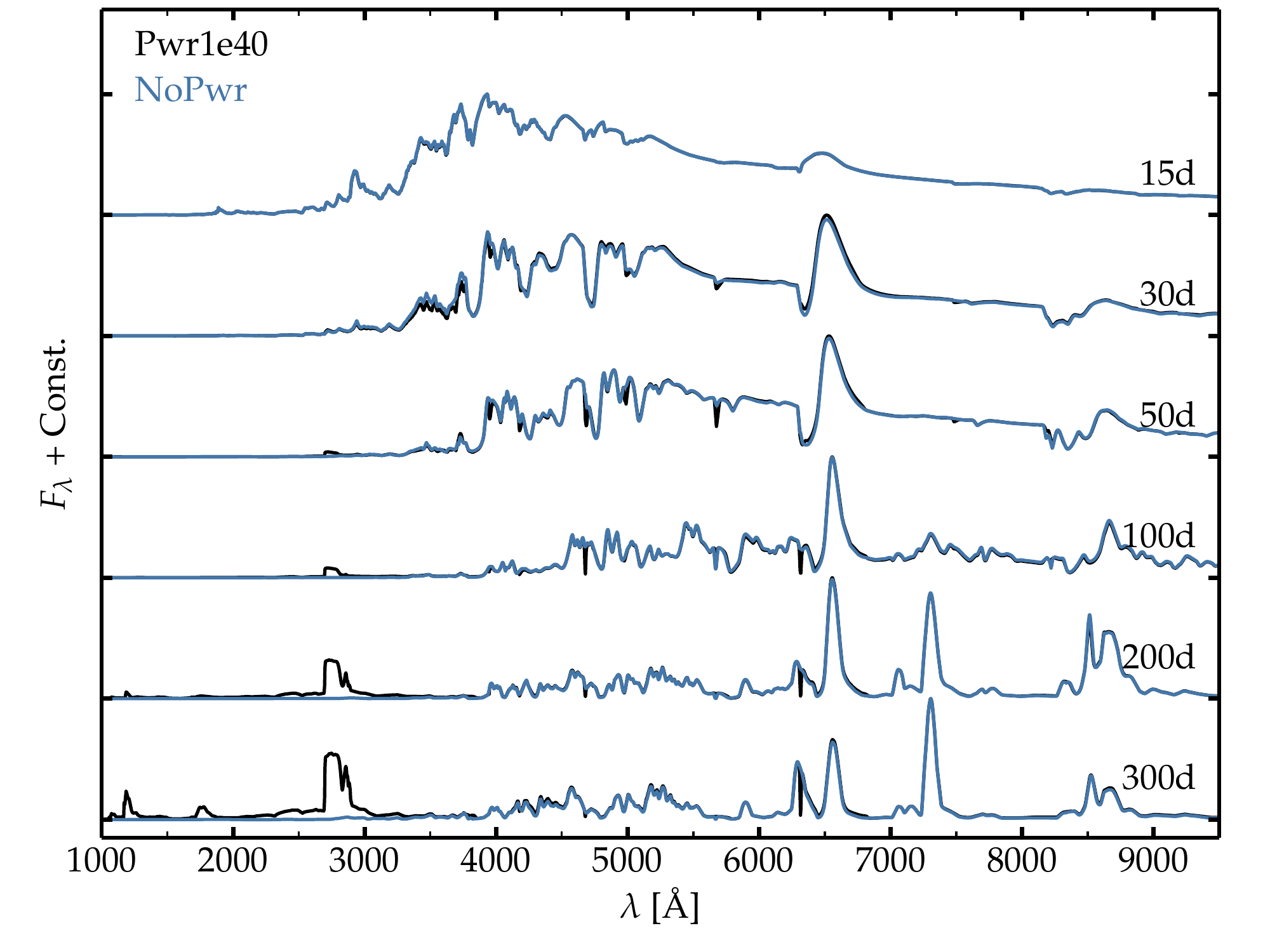}
\includegraphics[width=0.38\hsize]{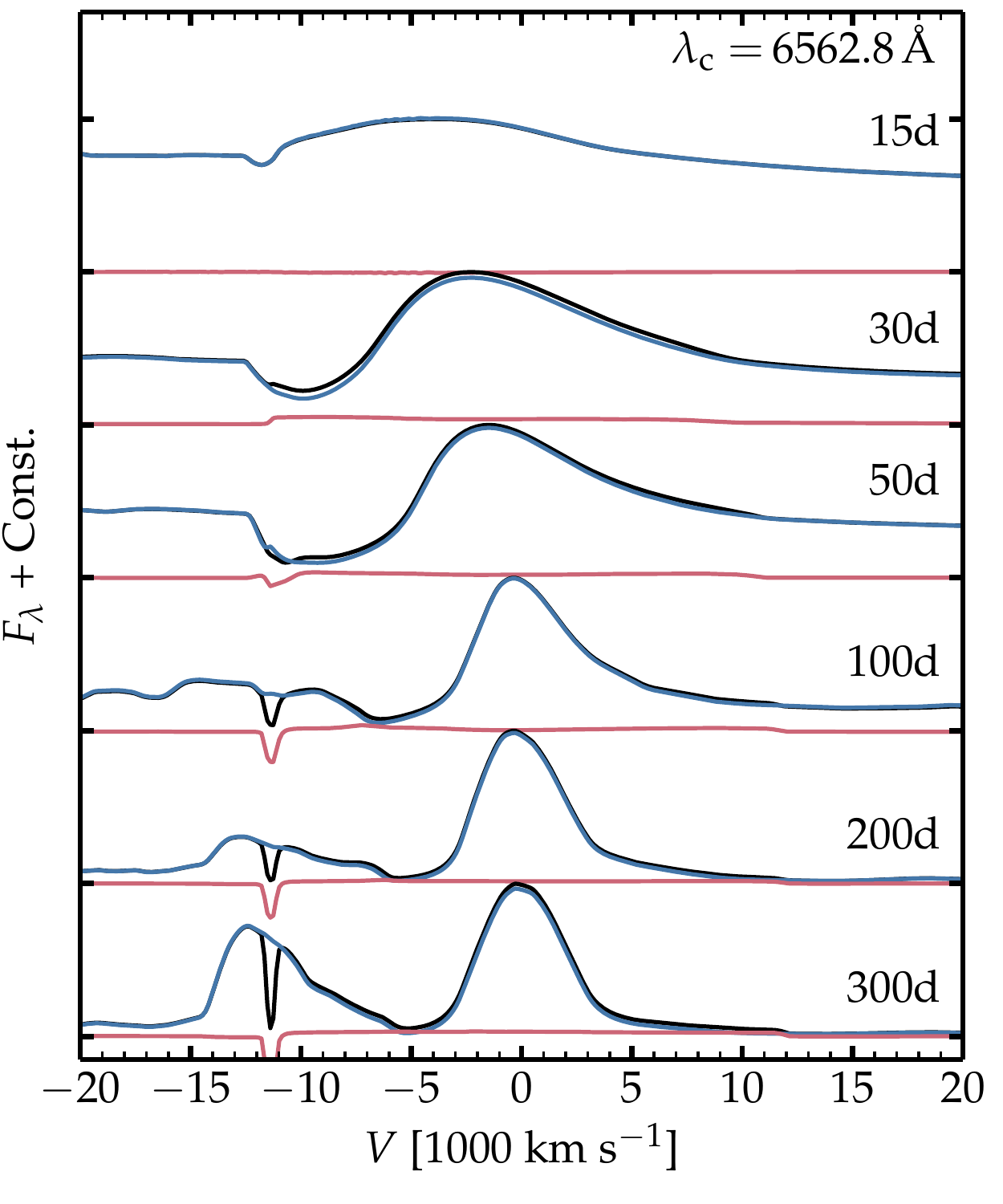}
\caption{Same as Fig.~\ref{fig_pwr2}, but now for the model Pwr1e40.
\label{fig_pwr1}
}
\end{figure*}

\begin{figure*}
\centering
\includegraphics[width=0.61\hsize]{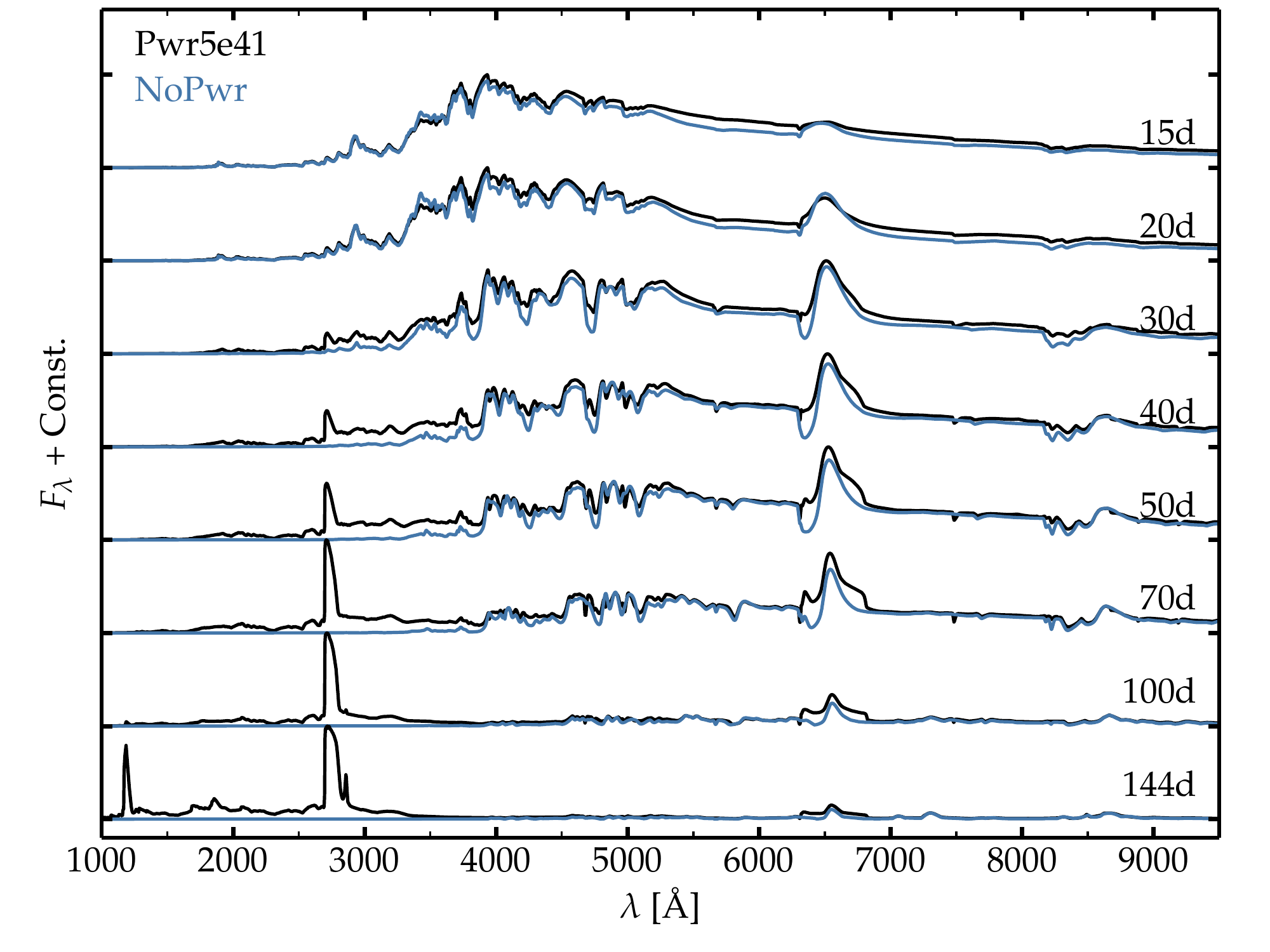}
\includegraphics[width=0.38\hsize]{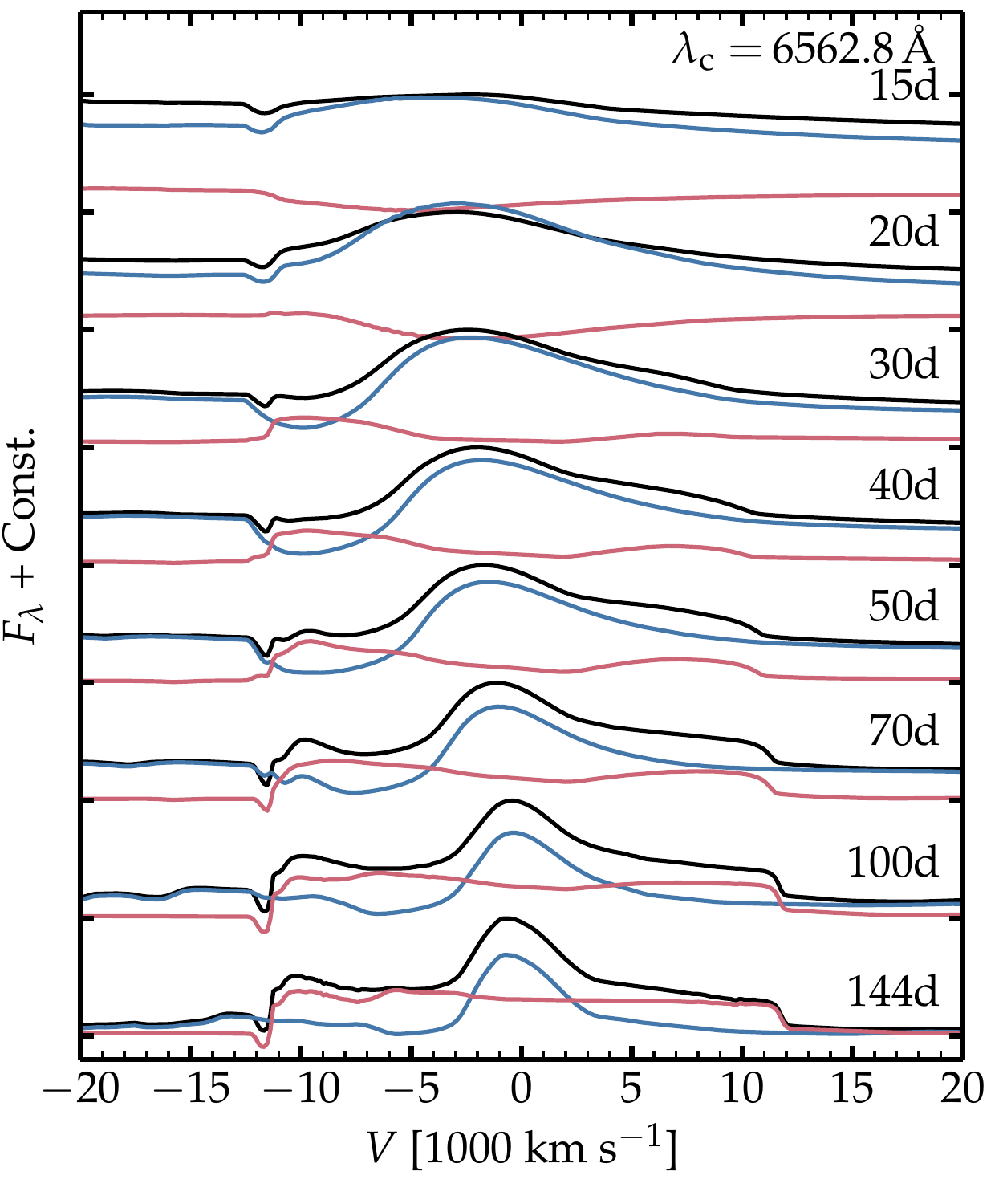}
\caption{Same as Fig.~\ref{fig_pwr2}, but now for the model Pwr5e41. The time sequence was stopped at 144\,d.
\label{fig_pwr3}
}
\end{figure*}

\begin{figure*}
\centering
\includegraphics[width=0.61\hsize]{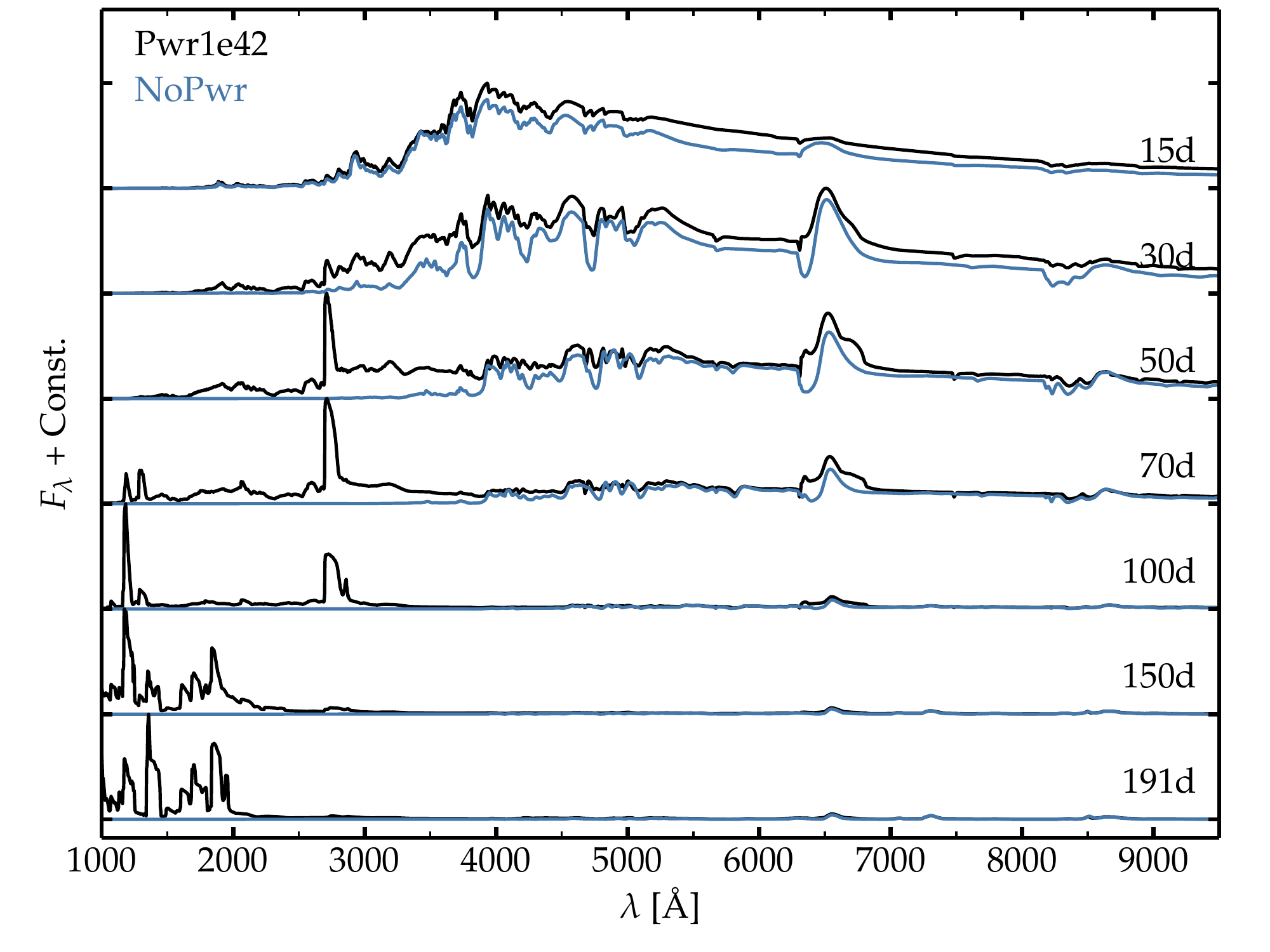}
\includegraphics[width=0.38\hsize]{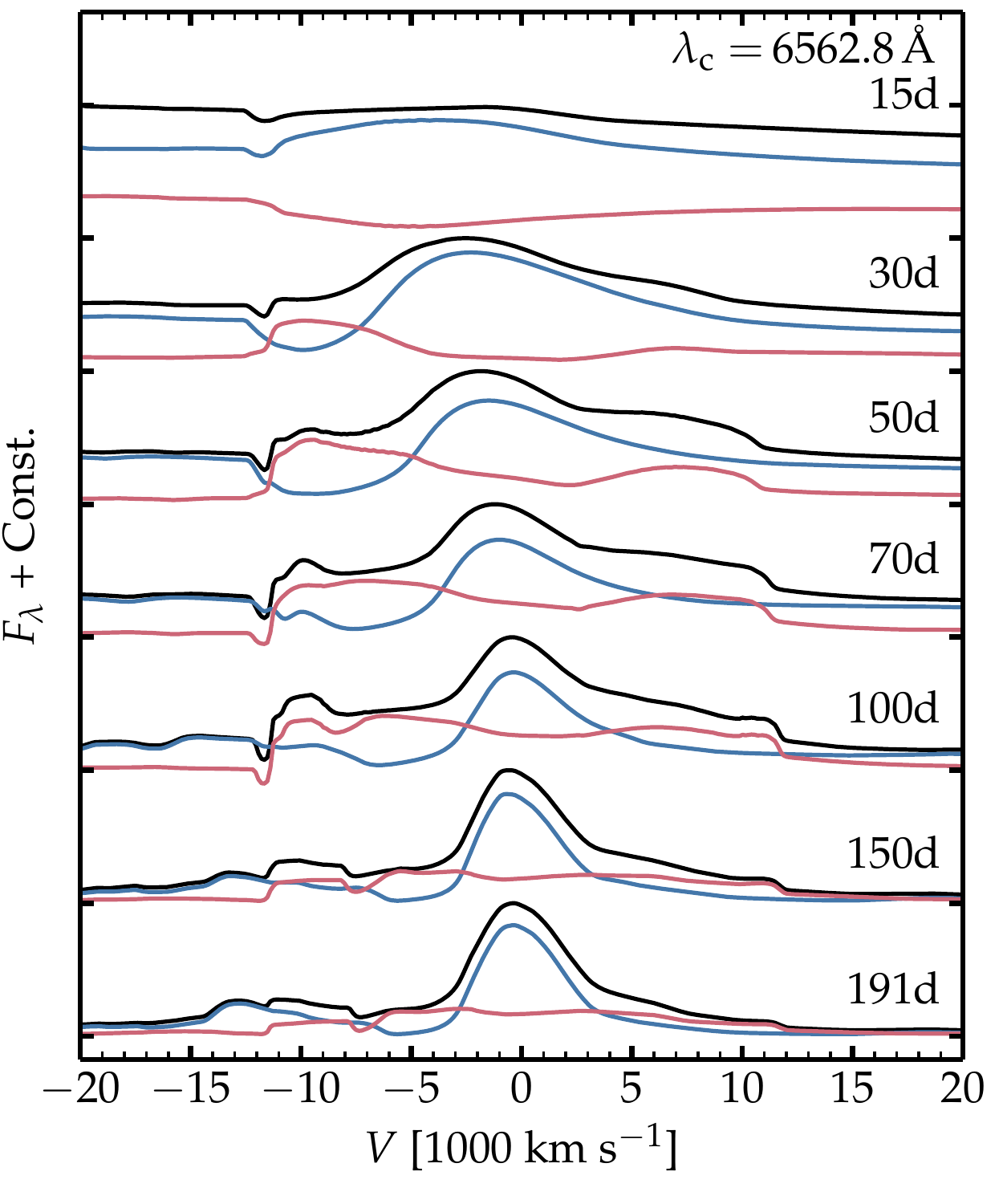}
\caption{Same as Fig.~\ref{fig_pwr2}, but now for the model Pwr1e42. The time sequence was stopped at 191\,d.
\label{fig_pwr4}
}
\end{figure*}

\begin{figure*}
\centering
\includegraphics[width=0.61\hsize]{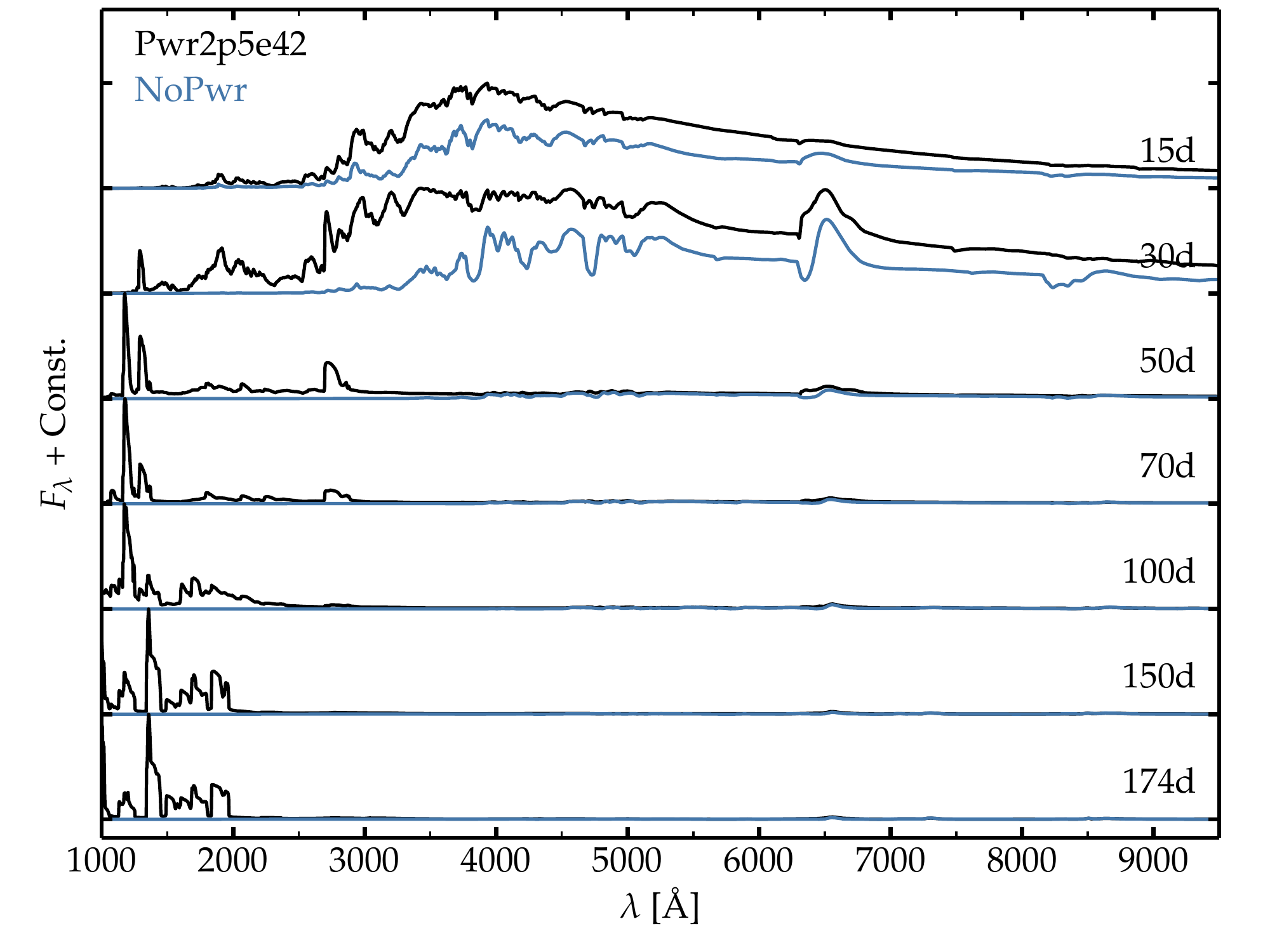}
\includegraphics[width=0.38\hsize]{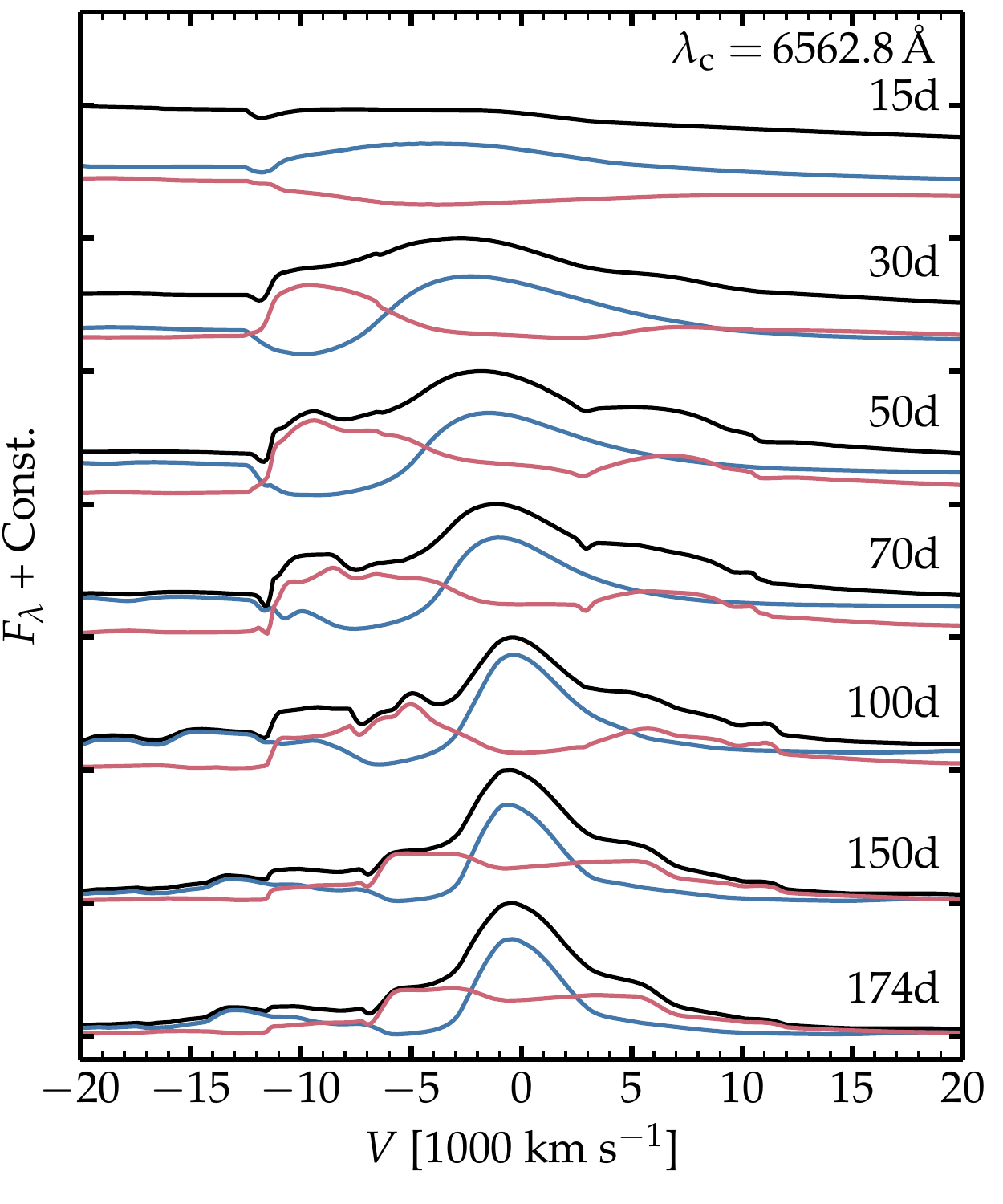}
\caption{Same as Fig.~\ref{fig_pwr2}, but now for the model Pwr2p5e42. The time sequence was stopped at 174\,d.
\label{fig_pwr4b}
}
\end{figure*}

\begin{figure*}
\centering
\includegraphics[width=0.61\hsize]{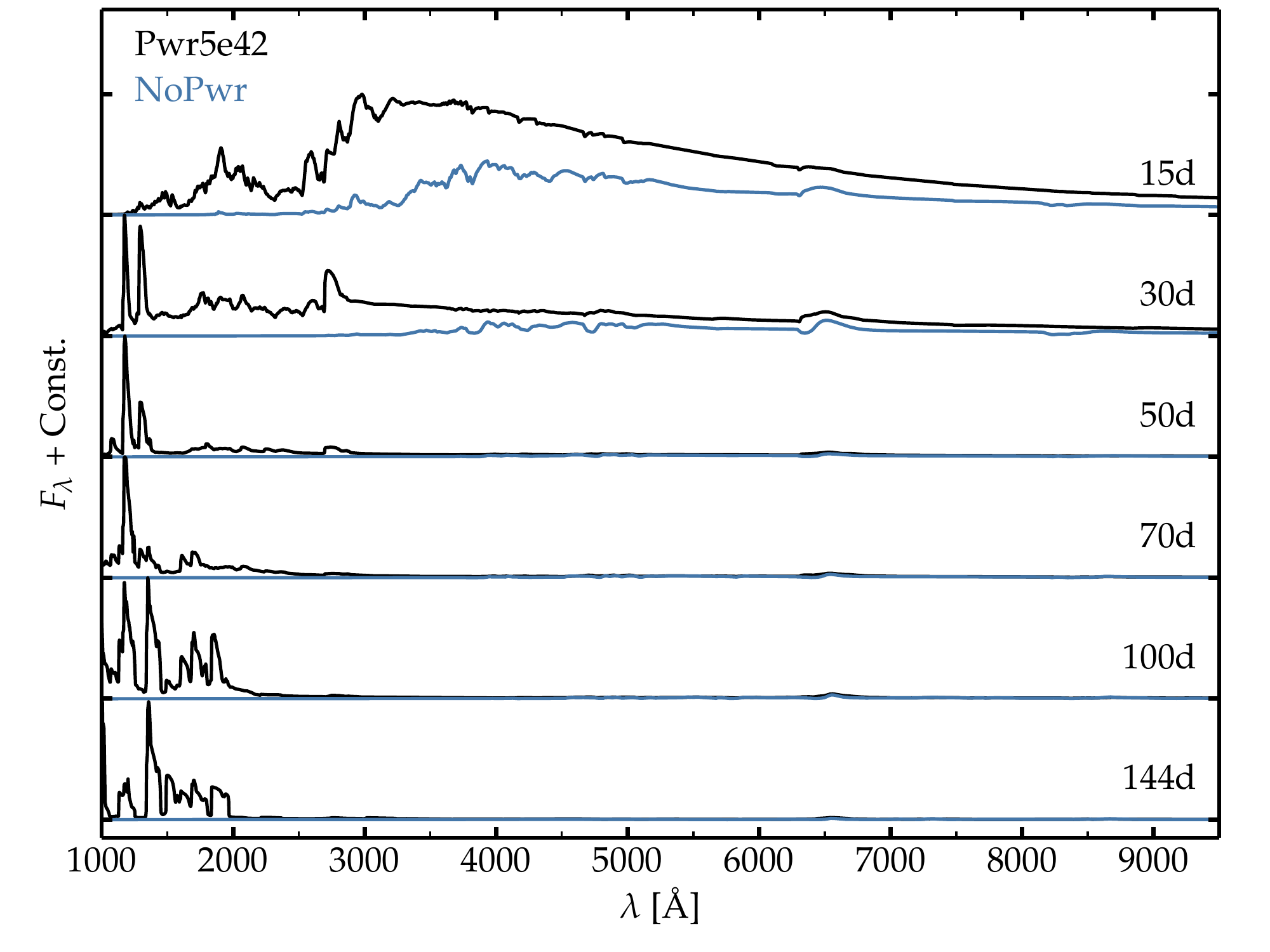}
\includegraphics[width=0.38\hsize]{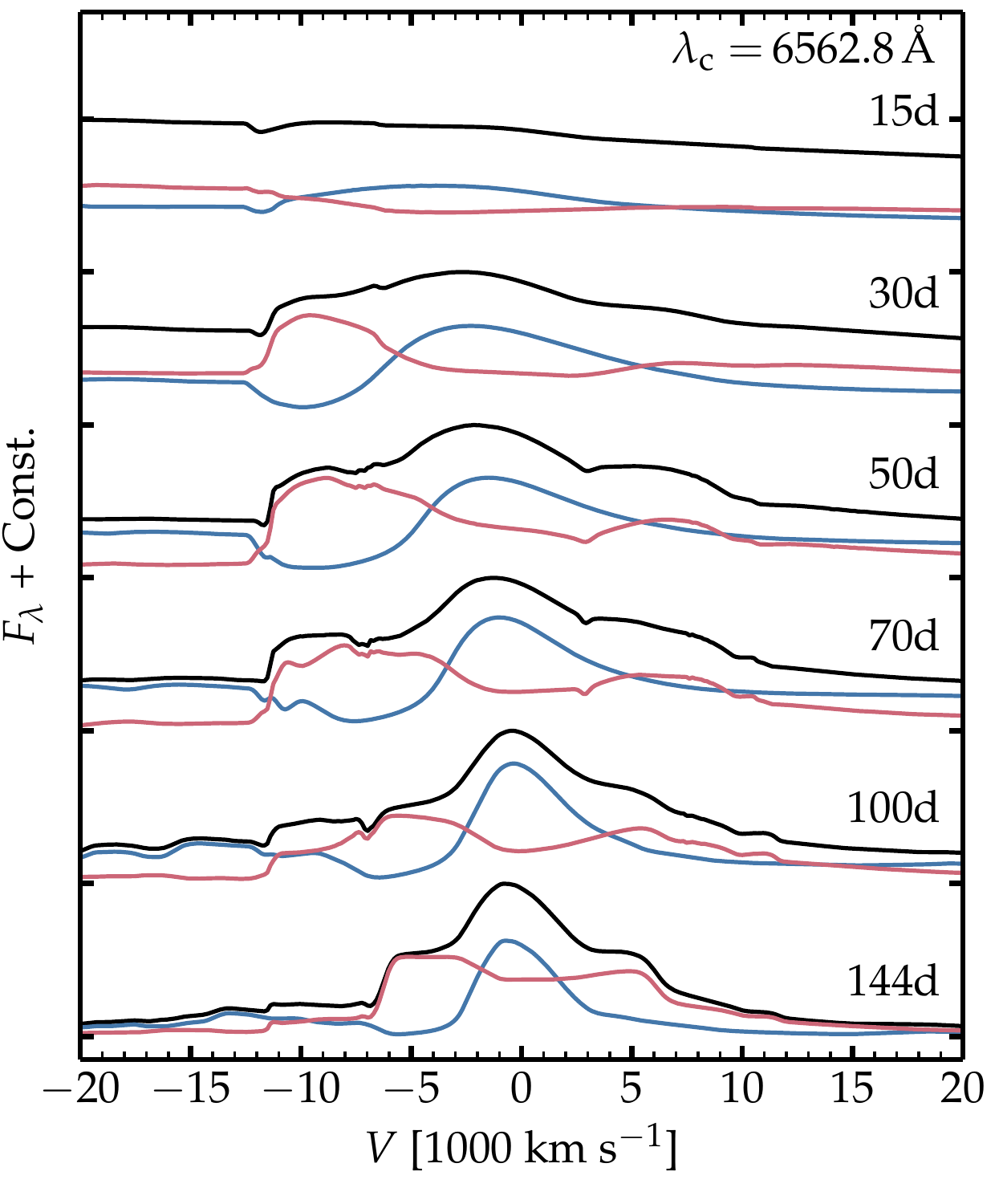}
\caption{Same as Fig.~\ref{fig_pwr2}, but now for the model Pwr5e42. The time sequence was stopped at 144\,d.
\label{fig_pwr5}
}
\end{figure*}

\begin{figure*}
\centering
\includegraphics[width=0.61\hsize]{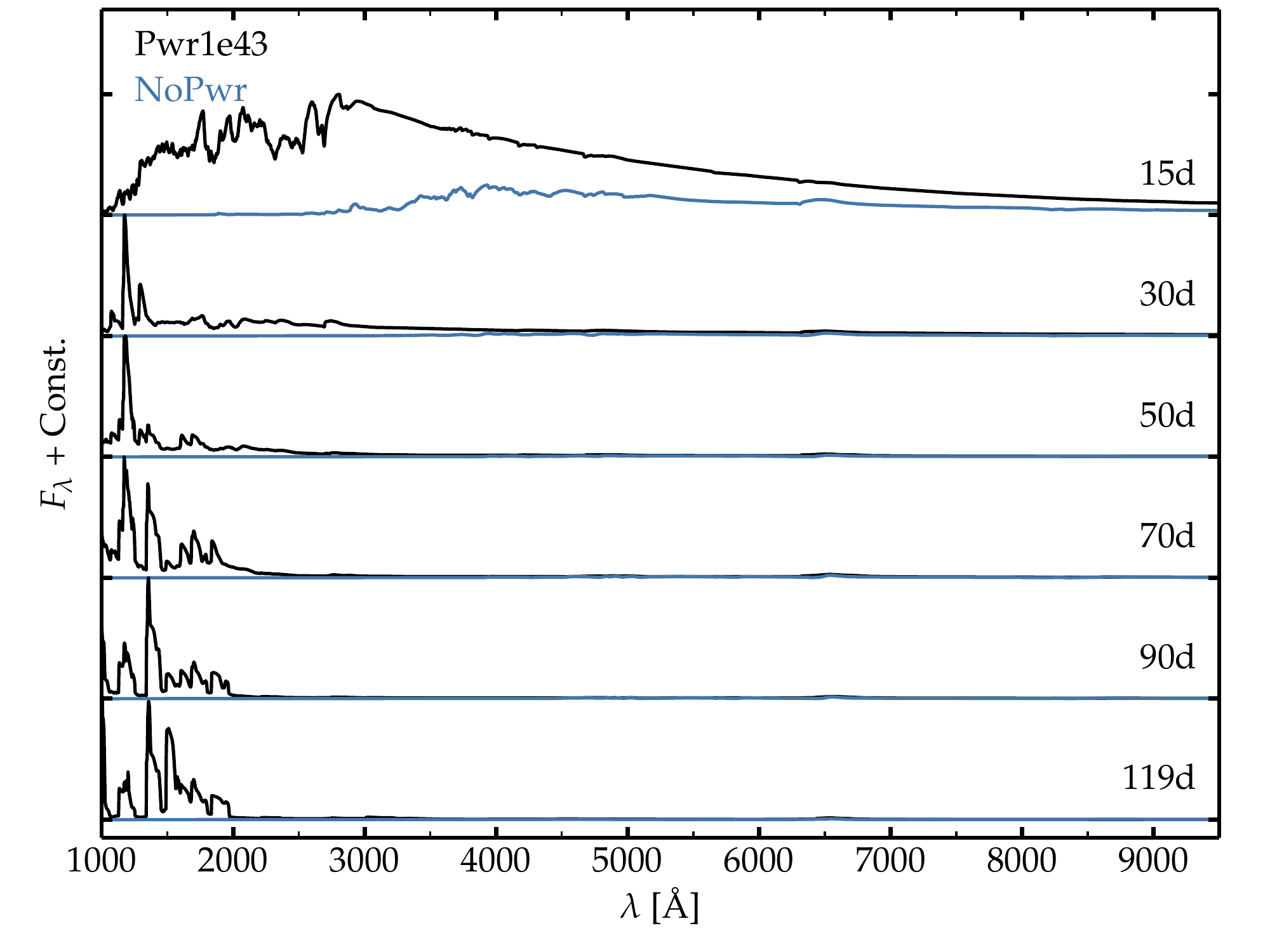}
\includegraphics[width=0.38\hsize]{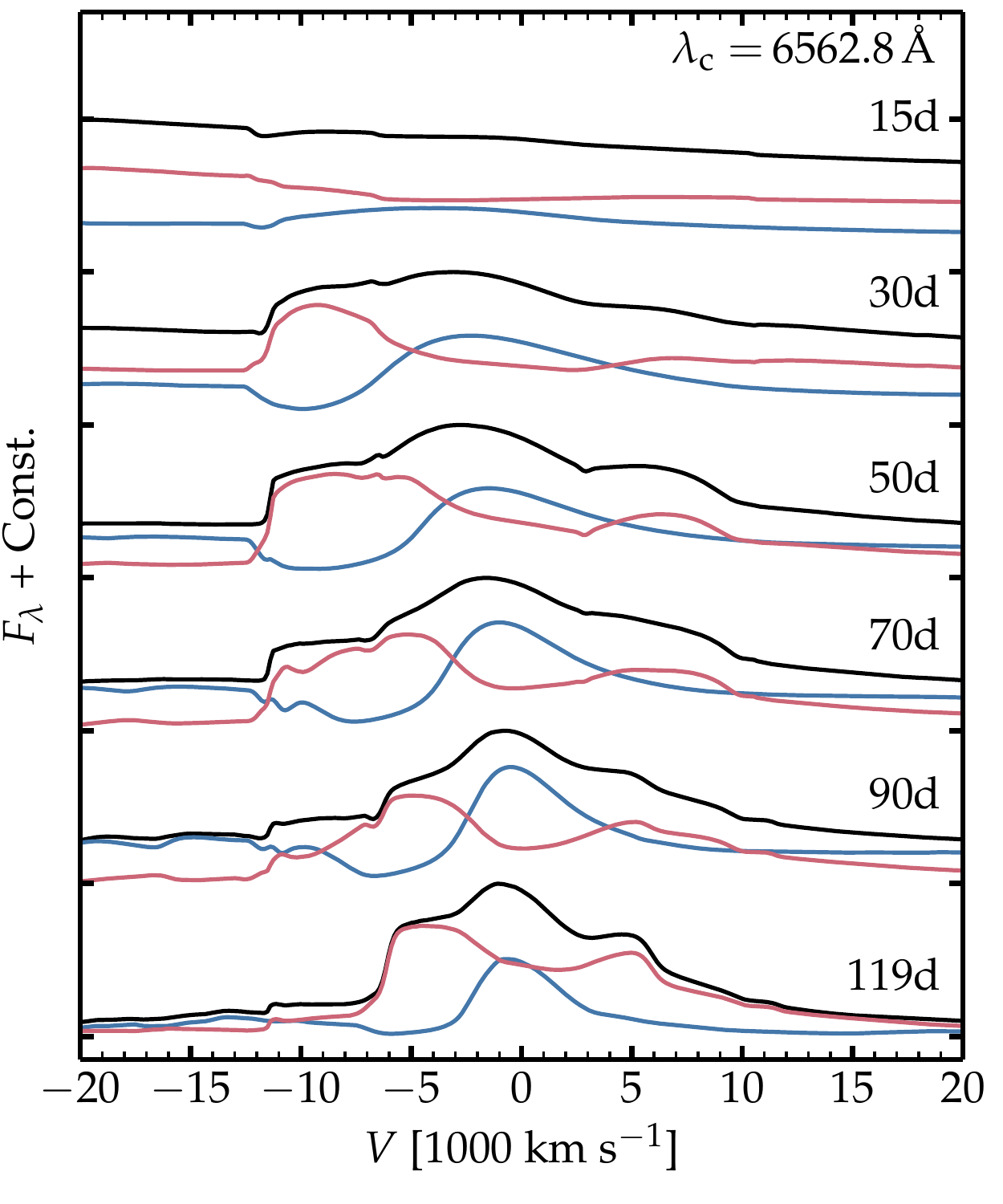}
\caption{Same as Fig.~\ref{fig_pwr2}, but now for the model Pwr1e43. The time sequence was stopped at 119\,d. Note, however, that the interaction already dominates the radiative contribution from the underlying ejecta such that, with the adopted constant interaction power, the subsequent spectral evolution should be slow and exhibit similar properties.
\label{fig_pwr6}
}
\end{figure*}

Figures~\ref{fig_pwr1} to \ref{fig_pwr6} are similar to Fig.~\ref{fig_pwr2}, but show the results for the whole set of models that cover from zero to the highest interaction power of 10$^{43}$\,\ergs. At high interaction power, the models were stopped earlier in the sequence, around 50\,d or at the onset of the nebular phase.

\end{document}